\documentstyle[12pt,epsfig]{article}     
\renewcommand{\theequation}{\thesection.\arabic{equation}}       
\newcommand{\be}{\begin{equation}}       
\newcommand{\ee}{\end{equation}}       
\newcommand{\bear}{\begin{eqnarray}}       
\newcommand{\eear}{\end{eqnarray}}       
\newcommand{\ba}{\begin{array}}       
\newcommand{\ea}{\end{array}}

\newcommand{\tg}{\tilde{g}}      
\newcommand{\ov}{\overline}      

\newskip\humongous \humongous=0pt plus 1000pt minus 1000pt

\newif\ifdtup

\def\oldreffmt#1{\rlap{[#1]} \hbox to 2\parindent{}}

\def\figfmt#1{\rlap{Figure {#1}} \hbox to 1in{}}       
       
\def\tr{\mathop{\rm tr}}

\def\VEV#1{\left\langle #1\right\rangle}

\def\slash#1{#1\!\!\!/\!\,\,}       
\def\beq{\begin{equation}}       
\def\eeq{\end{equation}}       
\def\bea{\begin{eqnarray}}       
\def\eea{\end{eqnarray}}       
\def\half{\frac{1}{2}}       
       
\def\bq{\begin{quote}}       
\def\eq{\end{quote}}

\def\half{\frac{1}{2}}            
\def \lta {\mathrel{\vcenter       
     {\hbox{$<$}\nointerlineskip\hbox{$\sim$}}}}       
        
\relax       

\newdimen\tdim       
\tdim=\unitlength       
\def\bar{\overline}

\begin{document}       
       
\pagestyle{empty}       
\begin{titlepage}       
\def\thepage {}    
       
\title{  \bf         
Dynamical Electroweak Breaking  \\    
and Latticized Extra Dimensions }        
\author{       
\bf Hsin-Chia Cheng$^1$\\[2mm]       
\bf  Christopher T. Hill$^2$ \\[2mm]       
\bf Jing Wang$^2$ \\ [2mm]       
{\small {\it $^1$Enrico Fermi Institute, The University of Chicago}}\\     
{\small {\it Chicago, Illinois, 60637 USA}}\\     
{\small {\it $^2$Fermi National Accelerator Laboratory}}\\     
{\small {\it P.O. Box 500, Batavia, Illinois 60510, USA}}     
\thanks{e-mail: hcheng@theory.uchicago.edu,     
hill@fnal.gov, jingw@fnal.gov }\\     
}     
\date{May, 2001}     
\maketitle     
\vspace*{-13.0cm}     
\noindent     
\begin{flushright}       
FERMILAB-Pub-01/079-T \\ [1mm]       
May, 2001       
\end{flushright}

\vspace*{14.1cm}       
\baselineskip=18pt       
       
\begin{abstract}       
       
  {\normalsize       
Using gauge invariant     
effective Lagrangians in $1+3$ dimensions      
describing the     
Standard Model in $1+4$ dimensions,     
we explore dynamical electroweak     
symmetry breaking. The Top Quark Seesaw     
model arises naturally, as     
well as the full CKM structure.  We include      
a discussion of effects of warping,      
and indicate how other dynamical schemes may also     
be realized.     
}      
\end{abstract}       
       
\vfill       
\end{titlepage}       
       
\baselineskip=18pt       
\renewcommand{\arraystretch}{1.5}     
\pagestyle{plain}       
\setcounter{page}{1}     
     
\section{Introduction}        
     
Recently we introduced the low energy effective     
Lagrangian in $1+3$ dimensions for the      
Standard Model  in a $D$ dimensional Yang-Mills gauge theory     
\cite{wang0}.      
Gauge fields, fermions,     
and Higgs scalars propagate in the bulk which     
is latticized \cite{wang1}.     
The extra dimensions, when     
described by the transverse lattice     
technique~\cite{trans}     
become a prescription for writing down an     
extension of the Standard Model in $1+3$ dimensions.     
As KK modes are discovered, they carry a hidden copy    
of the gauge group they represent    
through the hidden local symmetry of vector    
mesons \cite{hidden}.  Thus, the     
enlargement of the gauge group into the bulk is realized     
as one climbs the KK tower.  An independent approach,   
very similar to ours, was  
proposed in \cite{georgi}.

Our approach   
emphasizes the importance of a gauge invariant    
description of an infrared truncation of the theory.  There is  
significant utility in mapping the   
$D$-dimensional theory into an equivalent    
$1+3$ theory as a model building tool.      
We are lead to a chain of Standard Model gauge groups    
(this element of the scheme    
has a heritage, see e.g., \cite{earlier})    
and ``linking-Higgs     
fields'' which, in the broken phase,     
are the Wilson links in the extra dimensions, allowing    
hopping from one lattice brane to another~\cite{wang0,wang1}.    
These linking-Higgs fields  can      
be viewed as a valid    
UV description of the extra-dimensional theory up to the quartic    
Landau poles of our Higgs potential, where something like   
a superstring  phase transition probably occurs \cite{lykken}.

 We will    
henceforth    
refer to our approach of describing the $D$ theory with the    
$1+3$ dynamics as a {\em remodeled} extra dimensional theory.    
The {\em remodeling}, or latticization of compact extra dimensions    
to produce an effective $1+3$ Lagrangian with    
new dynamics, is in a sense the analogue of descending    
in supersymmetry from a full superspace action to an    
action in pure space-time.  Just as supersymmetry acts    
as an organizing principle and     
dictates constraints on the spacetime theory, so    
too an extra dimensional theory {\em remodeled} into $1+3$    
dictates a certain structure and dynamics.    
Moreover, we can map  the physical questions we wish to     
address, e.g.,  dynamics,     
topology, electroweak symmetry breaking, etc.,     
into conventional methods familiar to     
$1+3$ model builders.      
Everything is manifestly gauge invariant     
and renormalizable.     
Casting a given theory with new    
dynamics into  {\em remodeled } extra dimensions can yield     
insights and avenues for extension of the new dynamics.     
     
Our present task is to explore     
dynamical fermion bilinear condensate formation for the     
breaking of electroweak symmetry in the    
context of remodeled extra dimensions.      
A striking aspect of the Standard Model     
in the latticized bulk construction     
is that it provides the essential ingredients      
of a Topcolor model \cite{hill0}.     
Indeed, Topcolor is a dynamical gauge theory    
basis for top quark condensation    
\cite{topc} and involves rather uniquely  the imbedding     
of $SU(3)\rightarrow SU(3)_1\times SU(3)_2...$. Here     
the third generation feels the stronger $SU(3)_1$     
interaction, while the first and second generations     
feel the weaker $SU(3)_2$.   Such an imbedding, or     
enlargement of the $SU(3)$ gauge group is a natural     
consequence of extra dimensions with localized fermions     
\cite{wang0,wang1}. Indeed, Topcolor viewed    
as a remodeled extra dimensional theory anticipates    
the fermionic generations arising in    
a localized way in extra dimensions \cite{earlier,schmaltz}.     
    
Extra dimensional models with gauge fields    
in the bulk \cite{dienes}, or their remodeled counterparts,    
are inherently strongly coupled.    
We show that the inherent strong    
coupling expected in these models can     
naturally provide a dynamical     
condensation of $\VEV{\bar{t}t}$. In    
the remodeled description this is on a firm    
footing since the dynamics can be approximated reliably    
by a Nambu-Jona-Lasinio model. We should say that {\em a priori}    
nothing precludes the addition of more physics, e.g., supersymmetry    
or technicolor, etc.  We pursue Topcolor and Top Seesaw    
models at present    
because the remodeling of the $1+4$ Standard Model supplies all    
the ingredients for free!    
    
If we    
wanted to construct a pure Topcolor model, or a model    
such as Topcolor Assisted Technicolor \cite{hill0,lane}, we     
also  require a ``tilting'' mechanism     
to block the formation of a $\VEV{\bar{b}b}$.     
Again, the Standard Model in the latticized bulk     
provides the desired extra weak hypercharge    
imbedding $U(1)_Y\rightarrow     
U(1)_{Y1}\times U(1)_{Y2}...$ needed to     
tilt in the direction of the top condensate. The fact that     
the top--anti-top channel is the most attractive channel     
in a Standard Model configuration then drives the formation of     
the top condensate alone.     
     
In the present paper, however, we will explore a further aspect    
of the  dynamics of a remodeled $1+4$ theory with    
the Standard Model    
gauge structure propagating in the bulk.      
We will show that the Top Seesaw    
model \cite{topseesaw}, which may indeed be the best and most    
natural model of dynamical electroweak symmetry breaking,     
arises completely and naturally from extra dimensions.     
In a Top Seeaw model a top condensate forms    
with the natural electroweak mass gap, $\mu \sim 600$ GeV,     
but there exist    
additional vectorlike partners to the $t_R$ quark, usually designated    
by $\chi_R$ and $\chi_L$.      
These objects form heavier Dirac mass combinations    
as $M\bar{\chi}\chi$ and $m'\bar{\chi}_Lt_R$, and taken together the    
physical top mass is given by  $m_{top} = m'\mu/M$.      
The Top Seesaw affords a way    
to make a heavy top quark, and explain all of the electroweak breaking    
with a minimum amount of fine tuning.  It has a heavy Higgs    
boson $\sim 1$ TeV, yet is in full consistency with the $S-T$    
error ellipse constraints \cite{collins, he}.    
  Remarkably, the vectorlike $\chi$ quarks of  
 the Top Seesaw are also available for free    
from extra dimensions.  These are simply the ``roaming''    
$t_R$ quark in the bulk,  away from the domain wall that localizes     
it's chiral zero mode $t_R$.    
  
The possibility of generating    
top condensation (or other) schemes in the context    
of extra dimensions  has been developed     
previously in explicit continuum extra dimensions \cite{dobrescu};    
indeed, Dobrescu \cite{dobrescu2} first     
observed that dynamical electroweak symmetry    
breaking was a likely consequence    
of the strong coupling of QCD in extra dimensions.     
The geometric    
reasoning we inherit from extra dimensions  
leads us to a systematic way of extending    
the models.  Remodeled extra    
dimensions has led us in the present paper    
to the first theory of flavor physics from  
the Top Quark Seesaw, with CKM structure and light fermion    
masses.   We also show that one can readily  
construct a viable $4$th generation scheme along  
these lines. All fermions are condensed by the     
$SU(3)\times U(1)_Y$ structure on    
the $4$th generation-brane, and one can  
postulate a Majorana masses for the     
$\nu_{Ri}$ as well, allowing the   
Gell-Mann--Ramond--Slansky--Yanagida neutrino seesaw  
(see, e.g., \cite{luty} and references therein).  
    
Our present discussion will be largely schematic.     
We will describe the structure of the theory, and    
in a later work we will present the full    
phenomenology \cite{wang5}.    
To make the present discussion as transparent as possible     
we will ``thin the degrees of freedom.''     
Normally, we would approximate the bulk with a very large     
number of branes and interlinking Higgs fields.     
Presently, however, we will describe reduced     
$n$-brane models, in which $n$ is small,     
typically $n = 2, 3, 4, 5...$.  In our  minimal Top Seesaw    
scheme we have  $n=4$, i.e., there is one brane per generation    
and one extra spectator brane (required for technical reasons).    
Hence, in this case all of     
the bulk is approximated by a transverse lattice      
with four branes.  The gauge     
group we consider in $1+3$ dimensions for the $n$-brane model is     
$SU(3)^n\times SU(2)_L^n\times U(1)_Y^n$     
and we have $n-1$ link-Higgs fields per     
gauge group. Thus, we      
keep only the zero-modes and $n-1$ Kaluza-Klein (KK) modes     
for each gauge field.  We will also keep     
some of the vectorlike KK modes of the fermions,     
in particular for the third generation. The masses of     
the vectorlike KK fermions are controlled by     
the mechanism that produces the chiral     
fermions on the branes \cite{jackiw,kaplan} and these can be lifted     
to arbitrarily large Dirac masses, independent     
of the compactification scale.      
    
The thinning     
of degrees of freedom is a mathematical     
approximation to the full theory. It is presumably    
derived from the fine-grained theory by a Kadanoff-style    
renormalization group.  As a result, we expect many renormalization effects,   
and    
e.g., any translational invariance that may be    
softly broken by background fields of the short-distance theory    
can be lost in the thinned degrees of freedom of the effective theory.    
Our residual engineering freedom, leading to any given scheme,     
arises largely from the localization of the chiral     
fermions and the freedom to renormalize the    
linking-Higgs VEV's and gauge couplings in a non-translationally    
invariant way.  How all of this ultimately interfaces with    
flavor physics constraints, e.g., flavor changing neutral    
current constraints \cite{pomarol}, etc., remains to be examined in detail    
\cite{wang5}.    
    
Thus our models can be viewed    
as transverse lattice descriptions     
of a Standard Model     
in $1+4$ dimensions in     
which the gauge fields and fermions and Higgs all     
live in the bulk \cite{dienes,schmaltz,dobrescu}    
with thinned degrees of freedom.    
Alternatively, they are a new class of $1+3$ models    
with Topcolor \cite{hill0, topc} and     
Top Seesaw \cite{topseesaw} dynamics.    
The two pictures    
are equivalent through {\em remodeling}.

\section{Effective Lagrangians in      
Warped Latticized Backgrounds}     
    
We begin with some essential preliminaries on latticized    
extra dimensions.    
We wish to describe the low energy effective      
Lagrangian of, e.g., the Standard Model      
in $1+4$ dimensions using the transverse lattice,    
but we include presently effects that break translational    
invariance in $x^5$.        
We begin with the QCD content and allow a general     
background geometry described by a metric with     
dependence upon the extra dimension $x^5$.      
      
Consider the pure gauge Lagrangian in $1+3$ dimensions      
for $N+1$ copies of QCD:     
\be      
{\cal{L}}_{QCD}= -\sum_{j=0}^N\; \frac{1}{4\tg_{j}^2}       
G_{j\mu\nu}^B G_j^{B\mu\nu} +     
\sum_{j=1}^{N} D_{\mu}\Phi_j^\dagger D^{\mu}\Phi_j     
\ee     
in which we have $N+1$ gauge      
groups $SU(3)_j$ with gauge couplings $\tg_{j}$    
{\em that depend upon $j$} and $N$ link-Higgs fields,      
$\Phi_j$      
forming $(\bar{3}_{j-1},{3}_{j})$     
representations. The covariant derivative is      
defined as $D_{\mu}= \partial_{\mu} + i     
\sum_{j=0}^{N} A_{j\mu}^{B}T_j^{B}$.      
$T_j^{B}$ are the generators of     
the $i$th $SU(3)_i$ gauge symmetry, where $B$ is the color index.     
Thus, $[T_i,T_j]=0$ for $i\neq j$;      
$T_j^B$ annihilates a field that is singlet     
under the $SU(3)_j$; when the covariant derivative     
acts upon $\Phi_j$ we have a commutator of     
the gauge part with $\Phi_j$, $T^{B\dagger}_j$ acting on     
the left and $T_{j-1}^B$ acting on the right;     
the $j$th field strength is determined as    
usual, $G_{j\mu\nu}^{B} \propto     
\tr T_j^{B}[D_\mu,D_\nu]$, etc.      
     
We treat the $\Phi_j$ as explicit    
Higgs fields.    
Renormalizable potentials can be constructed for      
each of the link-Higgs fields,     
and we can always       
arrange the parameters in the potential such that the diagonal     
components of each $\Phi_j$ develop a common vacuum      
expectation value (VEV) $v_j$, while     
the Higgs and $U(1)$ pseudo-Nambu-Goldstone boson (PNGB)     
are arbitrarily heavy (for the perturbative    
unitarity constraint on this limit see ref. \cite{wang0}).     
Hence, each $\Phi_j$ becomes effectively a     
nonlinear-$\sigma$ model field \cite{wang0, wang1}:     
\be      
\Phi_j \rightarrow v_j\exp(i\phi^B_j T^B_j/ v).     
\ee     
  In our previous     
discussion ~\cite{wang0,wang1}, we assumed    
that $\tg_{j}$ and $v_j$ were      
common for all $N+1$ gauge groups and $N$ links,    
i.e., independent of $j$.     
This corresponds to      
a translationally invariant extra      
dimension with physical parameters independent of $x^5$.      
    
In general,     
we must consider non-uniform $\tg_{j}$ and $v_j$ in    
the remodeled theory.  These correspond to a large variety of    
possible effects. For example, we may have    
an extra dimension with non-trivial background metric and a space dependent     
gauge coupling. These effects can arise from a bulk cosmological constant,    
background    
space dependent dilaton field, or from other fields and the finite     
renormalization effects due to localization of    
these fields.  Alternatively,     
a background scalar field with nontrivial    
dependence upon $x^5$, $\varphi(x^5)$, and coupled to     
the gauge kinetic term,    
$(G^B_{\mu\nu})^2$, will give finite     
$x^5$ dependent renormalization of $\tg$.      
     
Let us consider presently the case of    
a warped geometry, where the metric will    
contain an overall warp-factor or background    
dilaton field, e.g., a Randall-Sundrum model \cite{randall}.    
The effect     
of the dilaton field can be seen through the implicit      
identification of     
the link-Higgs fields $\Phi_n$ with the Wilson lines:      
\be      
\Phi_j(x^\mu) = \exp \left(i \int_{(j-1)a}^{ja} dy A^B_5(x^\mu, y)       
T^B\right),     
\ee     
where $a$ is the lattice spacing.     
One finds:     
\be      
D_{\mu}\Phi_j^\dagger D^{\mu}\Phi_j \; \to \;     
\half a^2 v_j^2 G^B_{(j-\half)\mu 5} G^{B\mu 5}_{(j-\half)}.     
\ee     
Let us      
compare this with the $1+4$ dimensional action for the gauge field      
in the background metric:     
\be     
ds^2 = e^{-2\sigma(y)} \eta_{\mu\nu} dx^\mu dx^\nu- dy^2,      
\label{background}     
\ee     
We have for the gauge action:     
\bear     
{\cal L}_G &=&\int d^5 x \sqrt{g} g^{MN} g^{PQ} \frac{-1}{4g_5^2(y)}       
G^B_{MP} G^B_{NQ}       
\nonumber \\      
&=& \int d^4 x \int dy \;\frac{-1}{4g_5^2(y)} \left( 
\eta^{\lambda\kappa}      
G^B_{\mu\nu} G^{B\mu\nu} - 2e^{-2\sigma(y)} G^B_{\mu 5}      
G^{B\mu 5}\right),      
\eear      
where the indices $\mu,\,\nu$ are raised and lowered by the Minkowskian metric     
$\eta_{\mu\nu}$.     
We thus can see      
by comparison that the gauge coupling $\tg_j$ is related to the 5-dimensional     
gauge coupling by $\tg_j^2= g_5^2(ja)/a$, (assuming that $g_5$ is      
smoothly varying,)     
and $v_j$ is simply related to the warp factor by:     
\be     
\label{warp}      
\tg_{j-\half} v_j a = e^{-\sigma((j-\half)a)},\;     
\ee      
where      
\be      
\tg_{j-\half} \equiv \frac{g_5((j-\half)a)}{\sqrt{a}}.      
\ee      
For smoothly varying $\tg_j$ and $v_j$, we can make the following       
interpolation:      
\be      
\tg_{j-\half}^2 = \tg_{j-1}\, \tg_j, \quad \quad      
\tg_j^2\, v_j\, v_{j+1}\, a^2\, =\, e^{-2\sigma(ja)}\,\equiv\, e^{-2\sigma_j}.     
\ee      
An example with 3 lattice points is described in Appendix A.    
    
It is also straightforward to obtain     
the transverse lattice Lagrangian for scalar and fermion fields under the     
warped background metric. The action for a scalar field under the     
background (\ref{background}) is given by~\cite{Goldberger:1999wh}:     
\bear      
&&\int d^5 x \sqrt{g} \left( g^{MN} \partial_M H^\dagger \partial_N H       
 - m_H^2 H^\dagger H \right) \nonumber \\      
&=& \int d^4 x \int dy \left(e^{-2\sigma(y)}       
\partial_\mu H^\dagger \partial^\mu H      
- e^{-4\sigma(y)} \partial_5 H^\dagger \partial^5 H - e^{-4\sigma(y)}      
 m_H^2 H^\dagger H \right).      
\eear      
After discretization, we have      
\be     
\int d^4 x \sum_{j=0}^{N} \left(e^{-2\sigma_j} \partial_\mu H_j^\dagger      
\partial^\mu H_j     
- e^{-4\sigma_{j-\half}} \frac{1}{a^2} \left| H_j - \frac{\Phi_j}{v_j} H_{j-1}     
\right|^2 - e^{-4\sigma_j} m_{H}^2 H_j^\dagger H_j \right).     
\ee     
We can rescale $e^{-\sigma_j} H \to H$,       
the Lagrangian is then given by:     
\bear     
{\cal L}_S &=& \sum_{j=0}^{N} \Bigg\{\partial_\mu H_j^\dagger \partial^\mu H_j      
- \left( m_{H}^2 \, e^{-2\sigma_j} \,+\, \tg^2_{j-\half} \,v_j^2\,      
e^{2(\sigma_j-\sigma_{j-\half})}\right) |H_j|^2      
\\      
&-& \tg^2_{j-\half} v_j^2\, e^{-2(\sigma_{j-\half}-\sigma_{j-1})} \left|      
\frac{\Phi_j}{v_j} H_{j-1}\right|^2      
+ \bigg(\tg^2_{j-\half} v_j^2\,      
e^{(\sigma_j+\sigma_{j-1}-2\sigma_{j-\half})}      
H_j^\dagger \frac{\Phi_j}{v_j} H_{j-1} + h.c.\bigg) \Bigg\}.\nonumber      
\eear     
As discussed in the previous paper~\cite{wang0}, the aliphatic model     
corresponds to the $S^1/Z_2$ orbifold compactification of      
the extra dimension.     
The even field under $Z_2$ corresponds to the boundary condition     
$H_{-1}=H_0$, and the odd field under $Z_2$ corresponds to the boundary     
conditions $H_{-1}=H_{N}=0$. The mass parameter $m_H^2$ should be     
replaced by $m_{Hj}^2$ if it depends on $y$, which can come from      
a $y$-dependent VEV of some field or the renormalization effects.     
     
The action of a fermion under the background (\ref{background}) is     
given by~\cite{Grossman:2000ra,Chang:2000nh,Gherghetta:2000qt}:     
\be     
\int d^4 x \int dy\, e^{-\frac{3}{2}\sigma}\overline{\Psi} \,\left(     
i\gamma^\mu \partial_\mu -\gamma_5 e^{-\sigma} \partial_5     
-\half \gamma_5 (\partial_5 e^{-\sigma})\right)\,  e^{-\frac{3}{2}\sigma}\Psi     
-e^{-4\sigma}m_{\Psi}\, \overline{\Psi}\,     
 \Psi.     
\ee     
After rescaling and discretization, the fermion Lagrangian is given by     
\bear     
{\cal L}_F &=& \sum_{j=0}^N      
\Bigg\{\overline{\Psi}_j i\gamma^\mu \partial_\mu \Psi_j     
+\Bigg[      
\frac{e^{-\sigma_{j+\half}}}{a} \overline{\Psi}_{jR}     
\left(\frac{\Phi^\dagger_{j+1}}{v_{j+1}}\Psi_{(j+1)L} - \Psi_{jL} \right)     
+\, h.c. \Bigg]     
\nonumber \\     
&& -\frac{1}{2a} \left(e^{-\sigma_{j+\half}}-e^{-\sigma_{j-\half}}\right)     
(\overline{\Psi}_{jL}\Psi_{jR}-\overline{\Psi}_{jR}\Psi_{jL})     
- e^{-\sigma_j}m_{\Psi j}      
(\overline{\Psi}_{jL}\Psi_{jR}+\overline{\Psi}_{jR}\Psi_{jL}) \Bigg\}     
\nonumber \\     
&=& \sum_{j=0}^N \Bigg\{\overline{\Psi}_j i\gamma^\mu \partial_\mu \Psi_j     
+\bigg(\tg_{j-\half}\, v_j\,\overline{\Psi}_{jL} \,\frac{\Phi_j}{v_j} \,     
\Psi_{(j-1)R} + h.c.\bigg)     
\nonumber \\ &&      
-\bigg(\half \left(\tg_{j-\half} v_j +\tg_{j+\half} v_{j+1}\right) +      
e^{-\sigma_j}m_{\Psi j} \bigg)\, \overline{\Psi}_j \Psi_j \Bigg\},      
\label{warped_fermion}     
\eear     
where we have used the relation (\ref{warp}) and imposed the boundary      
conditions     
$\Psi_{-1R}=\Psi_{NR}=0$ and $\Psi_{(N+1)L}=\Psi_{NL}$, corresponding      
to having $\Psi_R \;(\Psi_L)$ odd (even) under $Z_2$. There is one more     
$\Psi_L$ than $\Psi_R$ at lattice $N$, so there is a massless left-handed      
chiral fermion     
left. The gauge anomaly must be canceled by including additional chiral     
fermions. Reversing the $Z_2$ parity of $\Psi_L$ and $\Psi_R$ gives rise a      
massless right-handed fermion. This can be obtained by imposing the boundary     
conditions $\Psi_{-1R}=\Psi_{0R}$, $\Psi_{0L}=\Psi_{(N+1)L}=0$. Alternatively,     
we can make the changes $L \leftrightarrow R,\; \tg \to -\tg,\; (a \to -a)$     
in the Lagrangian (\ref{warped_fermion}), (corresponding to an opposite     
sign for the Wilson term which is included to avoid the fermion doubling    
problem,)     
and impose the boundary conditions $\Psi_{-1L}=\Psi_{NL}=0,\;      
\Psi_{(N+1)R}=\Psi_{NR}$.

\section{Top Quark Seesaw from Remodeled Extra Dimensions}      
    
We consider a sequence of $n$-brane schemes.     
 We put one generation of fermions and a copy of      
$SU(2)_L\times U(1)_Y\times SU(3)$ on     
each brane. In addition, we have $n-1$ link-Higgs-fields (chiral fields,     
one for each gauge group). In the end we will have    
a set of links from a spectator brane to the up brane    
(which is defined as the brane on    
which the chiral up quark is localized),     
one set from up to charm, another     
from charm to top.  We will thus have constructed an ``aliphatic model,''    
as in \cite{wang0,wang1}.   
There are the usual zero-mode gauge     
fields and the $n-1$ KK modes, which are determined exactly.   
No Nambu-Goldstone boson (NGB) zero modes    
occur as is usually the case in Technicolor-like models;    
(indeed these models have nothing to do with Technicolor).     
     
We are  assuming throughout that we have an     
underlying Jackiw-Rebbi mechansim~\cite{jackiw} to trap     
the fermionic chiral modes at the specific    
locations in the bulk. This involves scalar fields    
in $1+4$, $\varphi(x^5)_q$, which couple to    
$\bar{\psi}_q\psi_q$ and have domain wall    
configurations on which chiral zero-mode solutions exist.    
Away from the domain wall the fermions are vectorlike    
and have large Dirac masses.    
[The  Jackiw-Rebbi mechanism on a discrete lattice   
is described in Appendix B.]      
For the remodeled description of matter fields,       
we exploit the fact that the chiral       
fermions can always be engineered on any    
given brane, with arbitrarily massive        
vectorlike KK modes partners      
on all branes, so we need keep only the chiral zero-modes    
and the lower mass vectorlike fermions.       
Indeed, it is an advantage of the     
remodeled $1+3$ formalism that we       
can do this;     
in a sense the chiral generations are put in by hand    
in the remodeled theory, and we retain only the     
minimal relevant    
information that defines the low energy effective Lagrangian.

We require a mechanism to make the        
bare $\tg_3$ coupling of $SU(3)_{C,j}$     
critically strong on the top brane    
$j$, such that the top quark will condense.       
Of course, with $N$ branes (of equal couplings) the bare $\tg_3$ coupling is already       
$\sqrt{N}$ times stronger than the QCD physical coupling.       
The freedom exists to choose    
an arbitrarily strong bare $\tg_3$ on brane $j$ for       
a variety of reasons as described in Section 2.      
For example, if the kink field       
that localizes the chiral fermions    
couples to       
$(G_{\mu\nu}^{B})^2$, it can give finite renormalization       
to the top brane gauge coupling constants    
and trigger the formation of the condensate    
(see below).  Any non-universal       
translational invariance breaking in $x^5$ may provide       
such a mechanism.       
       
The vectorlike fermions of the Top Seesaw arise     
in a simple way: they are the roaming $t_R$ (and/or    
$t_L$) in the bulk.  In a sense it is remarkable       
that all of the ingredients are present.       
In addition, we get Topflavor \cite{topflavor}, with the copies       
of the $SU(2)_L$ gauge groups.  Here arises a novel problem       
first noted in ref.~\cite{hill0}.  With large $SU(2)_L$        
couplings the instanton mediated baryon number violation       
mechanism of `t Hooft becomes potentially problematic.

Finally, we ask: how is CKM matrix generated?  We can put       
generational       
linking terms in by hand, which presumeably arise       
from an underlying mechanism of       
overlapping wave-functions for       
split fermions~\cite{schmaltz}.  In our remodeled formulation we get    
no more or less information out than is put in by    
localizing the fermions in the bulk in the first place.

\subsection{The Schematic Top Seesaw}      
    
Let us first briefly review the Top       
Seesaw model.  In a schematic form of the    
Top Seesaw model, QCD is embedded into the gauge groups      
$SU(3)_1 \times SU(3)_2$, with gauge couplings $\tg_{3,1}$ and $\tg_{3,2}$      
respectively. The relevant fermions transform under these gauge groups      
are (anomalies are dealt with by extension to include    
the $b$-quark) \cite{topseesaw}:    
\be      
T_L :({\bf 3, 1}),\quad \chi_R :({\bf 3, 1}),\quad t_R,\, \chi_L:      
({\bf 1, 3}),      
\ee      
where $T_L=(t_L, b_L)$ is the third generation left-handed $SU(2)_L$      
doublet, $\chi_L,\, \chi_R,\, t_R$ are $SU(2)_L$ singlet.      
We include a scalar field, $\Phi$, transforming as       
$({\bf 3, \bar{3}})$, and it develops a diagonal VEV,      
$\langle \Phi^i_j \rangle= v \delta^i_j$, which breaks the Topcolor      
to QCD,      
\be      
SU(3)_1 \times SU(3)_2 \to SU(3)_{QCD}.      
\ee      
The massive gauge bosons (colorons) have mass      
\be      
M^2 = (\tg_{3,1}^2 + \tg_{3,2}^2)\, v^2.      
\ee      
Since $\chi_L$, $t_R$ have the same gauge quantum numbers, we can write      
down an explicit Dirac mass term:      
\be      
\mu_{\chi t}\, \ov{\chi}_L t_R + h.c..      
\ee      
A second Dirac mass term between $\chi_L$ and $\chi_R$ can be induced from the      
Yukawa coupling to $\Phi$,      
\be      
\xi\, \ov{\chi}_R \Phi \chi_L + h.c.\; \to \mu_{\chi\chi}\,\ov{\chi}_R \chi_L      
+ h.c..      
\ee      
These masses are assume to be in the TeV range and have the       
order $\mu_{\chi t}<\mu_{\chi\chi}< M$.      
Below the scale $M$, various 4-fermion interactions are generated      
after integrating out the heavy gauge bosons. We assume that $\tg_{3,1}$      
is supercritical and $\gg \tg_{3,2}$. A $\ov{T}_L \chi_R$ condensate      
will form and break the electroweak symmetry.      
To obtain the correct electroweak      
breaking scale, $\ov{t}_L \chi_R$ should have a dynamical mass      
$m_{t\chi}\sim 600$ GeV~\cite{topc}.      
The mass matrix for the $t_{L,R},\;\chi_{L,R}$ is then  \cite{topseesaw}:    
\be      
(\ov{t}_L,\, \ov{\chi}_L)       
\left(      
\begin{array}{cc}      
0 & m_{t\chi} \\      
\mu_{\chi t} & \mu_{\chi\chi}      
\end{array}      
\right)      
\left(      
\begin{array}{c}      
t_R \\ \chi_R      
\end{array}      
\right)      
\ee      
The light eigenstate is identified as the top quark and have a mass      
\be      
m_t \sim m_{t\chi} \frac{\mu_{\chi t}}{\mu_{\chi\chi}}      
\ee      
The top quark mass is correctly produced for $\mu_{\chi\chi}/\mu_{\chi t}      
\sim 3.5$.  The model thus produces an acceptable dynamical    
electroweak symmetry breaking and a composite Higgs boson    
(composed of $\sim \bar{t}_L\chi_R$) with a fairly    
natural scale of the new physics (the QCD imbedding scale)    
of $\Lambda \sim $ few TeV.    
    
\subsection{Top Seesaw from Remodeled Extra Dimensions}     
    
\begin{figure}[t]    
\vspace{4cm}    
\includegraphics{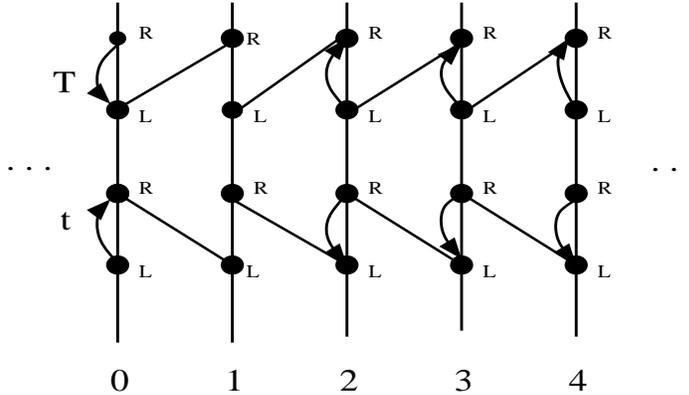}    
\vspace{1cm}    
\caption{A left-handed chiral zero mode $T_L$ (right-handed    
$t_R$) is    
localized on brane 1, by coupling to a kink in a    
background field $\varphi(x^5)$ which gives a   
negative (positive) Dirac mass to    
the right of brane 1 and negative (positive)  
to the left of brane 1.  We denote the negative (positive)  
Dirac mass by the up-arrow (down-arrow) curved links    
on each brane.  The trapping Dirac mass,  
with a coarse grain lattice, can alternatively be  
added to link terms on one side of the zero-mode   
as in Appendix B.  We use the definition of
the derivative for $T_L$ with 
linking Higgs fields   
with $L,j-1$ hopping to $R,j$, represented    
by the diagonal links between nearest neighbor branes,  
and $\sim -\bar{T}_{R,j}T_{L,j}$  
are vertical links on a given brane eq.(B.2). 
For $t_R$ we use the definition eq.(B.1).   
We keep only the lowest lying vectorlike modes in the picture.    
}   
\label{first}     
\end{figure}

All of the ingredients  of a Topcolor    
scenario, in particular the Top Qaurk seesaw,    
are present in an extra-dimensional scheme.    
We assume that we have only    
the fermions $T$ and $t$ in    
$1+4$ dimensions. The $\chi$ fields will appear automatically    
as the vectorlike KK mode components of these    
fields.    
The fermions are coupled in $1+4$ to    
a background field as $\varphi(x^5)\bar{T}T$    
and $\varphi(x^5)\bar{t}t$ and we assume that    
$\varphi(x^5)$ produces a domain wall kink    
at $x^5_0$ which we identify in our     
latticized approximation as the brane 1 in the Figures.  
Before the formation of the top condensate the   
top quark configuration on the  
lattice branes is depicted in Fig.[\ref{first}].

\begin{figure}[t]    
\vspace{4cm}    
\includegraphics{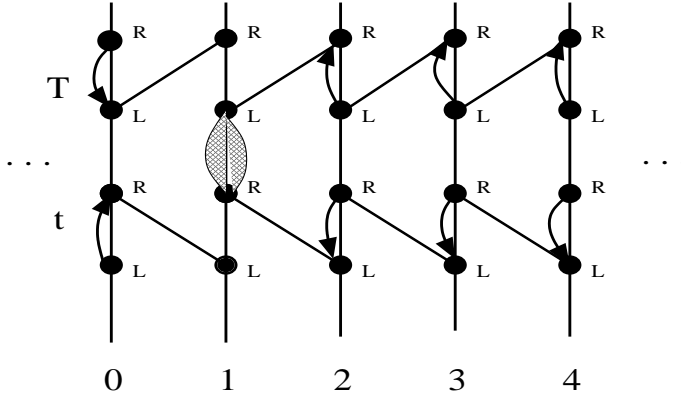}    
\vspace{1cm}    
\caption{A condensate  $\VEV{\bar{T}_Lt_R}$ forms on brane 1 when    
the $SU(3)_1$ coupling constant $\tg_{3,1}$ is    
supercritical. This can be triggered from $\varphi(x^5)(G_{\mu\nu}^2)$     
in the $1+4$ underlying theory, but is a free parameter choice in     
the $1+3$ effective Lagrangian.}    
\label{topcondensate}     
\end{figure}

The basic idea underlying the formation of a condensate    
is to allow a particular gauge coupling constant to become supercritical    
on a particular brane.  In Fig.~\ref{topcondensate} we show the formation    
of the condensate $\VEV{\bar{T}_Lt_R}$ on brane 1 where    
we assume that the $SU(3)_1$ coupling constant, $\tg_{3,1}$ is    
supercritical, i.e., in the NJL model approximation    
to QCD $3\tg^2_{3,1}/8\pi^2 >1$.     
    
A trigger mechanism in the $1+4$ theory    
for supercritical     
coupling at the location of the trapping    
domain wall can arise from a coupling of     
$\varphi$ to the squared field strength $(G_{\mu\nu}^a)^2$    
such that the gauge Lagrangian in $1+4$ becomes:    
\be    
\left[ -\frac{1}{4g_3^2 } - \frac{\lambda \varphi^2}{4M^2} \right] (G_{\mu\nu}^a)^2    
\ee    
Such a coupling will always be induced by the fermion fields    
which couple to the gauge fields.    
We assume $\varphi\rightarrow M$ ($\varphi\rightarrow -M$)    
for $x^5\rightarrow R$ ($x^5\rightarrow 0$).  For $\lambda>0$    
the action is well-behaved, and off the domain wall the    
effective coupling constant, $\bar{g}_3^2 = {g}_3^2/(1+\lambda)$    
is suppressed. On the domain wall the effective coupling    
is then ${g}_3^2$ which we assume is supercritical.    
Moreover, the condensate is generally suppressed,    
for fixed coupling ${g}_3^2$ in NJL     
model approximation for fields with large Dirac masses, so    
we expect only the chiral fields to pair up. In fact, one    
need not appeal to the trigger mechanism alluded to above, but    
it is a useful way to suppress $\bar{g}_3^2$ elsewhere in    
the bulk, and such operators are expected on general grounds when    
we construct the renormalized effective Lagrangian with    
fewer degrees of freedom.     
    
In our latticized $1+3$ decription the     
varying coupling constants $\tg_{3,j}$ and Dirac    
mass terms can be put in ``by hand'' as defining parameters.

\begin{figure}[t]    
\vspace{5cm}    
\includegraphics{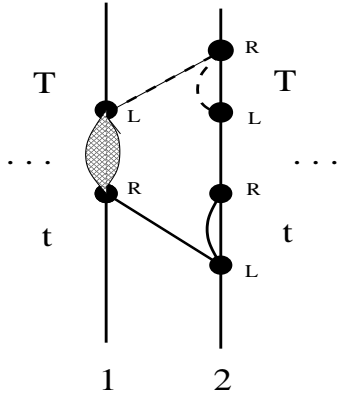}    
\vspace{1cm}    
\caption{Two brane approximation. In the limit    
that  $T_2$ decouples this is just the    
original Top Seesaw Model of \cite{topseesaw}.}      
\label{topseesaw}    
\end{figure}    
    
To derive the Top Seesaw Model from extra dimensions, we thin the      
degrees of freedom of the extra dimension to $2$-branes.      
There is an $SU(3)_j$ on each brane. The scalar field $\Phi$ which breaks      
$SU(3)_1\times SU(3)_2$ down to $SU(3)_{QCD}$ is     
now just the link-Higgs field and has exactly the right    
structure for Topcolor breaking.      
The $SU(2)_L$ doublet and singlet quark propagate in    
the $1+4$ bulk, and in the latticized scheme      
are represented by the fields $T_j,\,t_j,\, j=1,2$ on the two branes:      
\be      
\begin{array}{cccc}      
\multicolumn{2}{c}{SU(3)_1} & \qquad\qquad SU(3)_2 & \\      
T_{1L}  && T_{2R} & T_{2L} \\      
t_{1R} && t_{2L} & t_{2R} \\      
\end{array}      
\ee      
We have projected out the chiral partners    
$T_{1R}$ and $t_{1L}$ by   
coupling to the background localizing field with  
a domain wall kink at brane 1, which    
produces  chiral fermions.     
The kinetic terms in the extra dimension  give rise      
the mass terms in the $1+3$ effective Lagrangian  
that interconnect $T_{1L}$ to $T_{2R}$ etc.  
The backgound localizing fields $\varphi$ also produces  
the Dirac masses that interlink, e.g. $T_{2R}$   
and $T_{2L}$, etc.  So, in the two brane  
approximation we have:    
\be       
m_{T_{22}}\ov{T}_{2L} T_{2R}+     
m_{T_{12}}\ov{T}_{1L} \frac{\Phi^\dagger}{v} T_{2R}+    
m_{t_{22}}\ov{t}_{2L} t_{2R}+    
m_{t_{12}}\ov{t}_{1R} \frac{\Phi^\dagger}{v} t_{2L}+h.c.,      
\ee      
with      
\be      
m_{T_{12}} = -m_{t_{12}} \approx -\sqrt{\tg_{3,1}\tg_{3,2}}\, v      
\qquad    
m_{T_{22}} \sim \tilde{g}_{3,2}v + h_T\varphi,\;\; m_{t_{22}} \sim      
\tilde{g}_{3,2}v + h_t\varphi \qquad  h_T >> h_t   
\ee      
This configuration is shown in Fig.[\ref{topseesaw}].    
  
We can see explicitly that this matches onto  
the schematic Top Seesaw model.    
To match we first assume that the     
$\varphi$ contribution to the $\ov{T}_{2L} T_{2R}$ mass      
term is so large that      
${T}_{2L},\, T_{2R}$ decouple. Then, the $2$-brane      
model is identical to the schematic Top Seesaw model described above     
through the  following identification:     
\bear      
&T_L= T_{1L},\quad \chi_R= t_{1R},\quad      
\chi_L=t_{2L}, \quad      
t_R=t_{2R}, \nonumber \\      
&\mu_{\chi t}= m_{t_{22}}, \quad \mu_{\chi\chi}= m_{t_{12}}.      
\eear       
For a supercritical gauge coupling  $\tg_{3,1} $,      
the $\langle\ov{T}_{1L} t_{1R}\rangle$ condensate will form,       
breaking the electroweak symmetry. The top quark mass is     
then obtained from the seesaw mechanism.

\subsection{The Light Generations and Flavor Physics}      
      
We now consider all three fermionic    
generations of the Standard Model in     
the latticized bulk.     
We discuss the issue of    
how we can generate light quark masses and mixings     
in a generalized geometric Top Seesaw scenario.     
    
Clearly, in order to generate light      
fermion masses from the third generation condensates,      
some flavor mixing terms must be present.  Small masses and mixings     
can be generated in $1+4$ models    
by the overlap of the Higgs and fermion     
wavefunctions in the extra dimension~\cite{schmaltz}     
and/or small flavor mixing effects    
arising from localization.     
We examine this mechanism in the latticized extra dimension with     
the simplest flavor mixing mass terms. We find that the light     
generation fermion masses are generated radiatively in this picture.     
     
There is a copy     
of the $SU(3)\times SU(2)_L\times U(1)_Y$ Standard    
Model gauge group on each brane,     
with gauge couplings $\tg_{a,j}$ respectively, where     
$a=1, 2, 3,$ is the gauge group index and $j=0, 1, 2, 3$ is the brane index.     
There are link fields $\Phi_{a,j},\, a=1,2,3;\, j=1,2, 3$ which break     
the full     
$SU(3)^4\times SU(2)^4_L\times U(1)^4_Y$ gauge group     
down to the Standard Model $SU(3)\times SU(2)_L\times U(1)_Y$.     
    
We will denote the     
3 generation $SU(2)_L$ doublet quarks with uppercase letters ($T,\, C,\,     
U$) and $SU(2)_L$ singlet fermions with lower case letters ($t,\, c,\, u$)     
respectively.     
We assume that the third generation fermions propagates on all branes,     
with the localization removing the right-handed $SU(2)_L$ doublets     
and the left-handed singlets on brane $0$.  Hence the third generation    
fields $T_j$ and $t_j$, etc., carry the brane index $j$, while    
the  $C,c$  ($U,u$) are localized on brane $1$ ($2$). The localization    
of the top, charm and up quarks is accomplished with additional    
$\varphi(x^5)_t$,  $\varphi(x^5)_c$ and $\varphi(x^5)_u$ fields that produce    
domain walls in the underlying $1+4$ theory.    
    
If we assume that only the  $\tg_{3,0}$     
$SU(3)$ coupling constant is supercritical,      
this then drives the    
formation of the condensate $\VEV{\ov{T}_{L0}     
t_{R0}}$, breaking the electroweak symmetry.  The top     
quark mass is then obtained from the generalized seesaw mechanism.     
In general     
the left-- and the right--handed top quark zero modes are linear     
combinations of $T_{Lj}$ and $t_{Rj}$ (and $C_L,\, U_L,\, c_R,\, u_R$     
after including flavor mixings).     
\be     
T_L^{(0)}= \sum_{j=0}^3 \alpha_{T_j} T_{Lj}\; (+ \alpha_{C} C_L +     
\alpha_{U} U_L),\quad      
t_R^{(0)}= \sum_{j=0}^3 \alpha_{t_j} t_{Rj}\; (+ \alpha_c c_R +     
\alpha_u u_R),     
\ee     
where $\alpha_{T_j},\, \alpha_{t_j}$ are coefficients determined by the     
direct and link mass terms among $T_{L,R}$'s and $t_{L,R}$'s.     
The top quark mass is suppressed by the mixings $\alpha_{T_0}$     
and $\alpha_{t_0}$,     
\be     
m_t \sim \alpha_{T_0}\alpha_{t_0} \times 600\; {\rm GeV}.     
\ee     
     
\begin{figure}[t]    
\vspace{5cm}    
\includegraphics{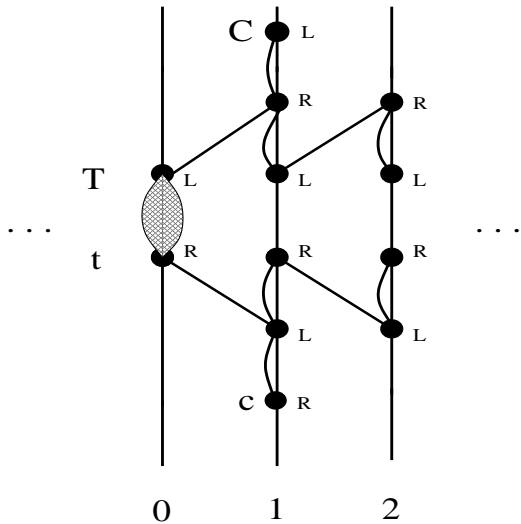}    
\vspace{1cm}    
\caption{Three brane approximation    
incorporating charm, where $C =(c,s)_L$ is a doublet    
zero-mode,    
and $c=c_R$ is a singlet zero mode, both trapped on brane    
1 (we assume the vectorlike partners of $C$ and $c$    
are decoupled).  The Dirac flavor mixing between    
$\bar{C}_L T_{R1}$ and $\bar{c}_Rt_{L1}$ can be rotated    
away by redefinitions of $T_{R1}$ and $t_{L1}$.     
}      
\label{charm1}    
\end{figure}

Let us first thin the degrees of freedom of the extra dimension to a      
$3$-brane model, and we consider first the generation    
of the charm quark mass.      
This configuration    
is as shown in Fig.[\ref{charm1}].

To generate the charm quark mass, we include flavor-mixing     
mass terms.  In the underlying $1+4$ theory we    
might suppose that these can arise    
on a given brane from couplings of the    
form, e.g.,      
$\epsilon \varphi(x^5)_t\bar{C}_L\psi_R $.    
In the $1+3$ theory this is    
a common Dirac mass    
on brane~1 that mixes all    
fermions with equivalent quantum numbers.    
However,     
the direct contact mass terms     
$\ov{C}_L T_{R1},\, \ov{t}_{L1} c_R,$     
can all be rotated away by redefinitions    
of the fields $T_{R1}$ and ${t}_{L1}$.       
This can be seen by considering the overall mass term     
on brane~1,   
$ m'\ov{t}_{L1} c_R + M \ov{t}_{L1}t_{R1}$, where the second term    
is just the mass of the Dirac (non-chiral) vectorlike $t$ quark    
on brane~1. Thus, by redefining $c_{R} \rightarrow (-m' t_{R1}    
+Mc_R)/\sqrt{M^2 + m'^2}$ and $t_{R1} \rightarrow (M t_{R1}    
+m' c_R)/\sqrt{M^2 + m'^2}$    
we eliminate the direct charm quark mass term.     
The important point is that this    
redefinition involves only fields on the common brane~1,    
and there is no residual kinetic term mixing since all of the    
kinetic terms involve the same gauge fields $A^B_{\mu,1}$.    
In order to generate a surviving mass term for the charm quark    
we thus need additional terms to frustrate the chiral redefinition.     
Such terms are seen to be present when we consider brane 2.    
    
\begin{figure}[t]    
\vspace{5cm}    
\includegraphics{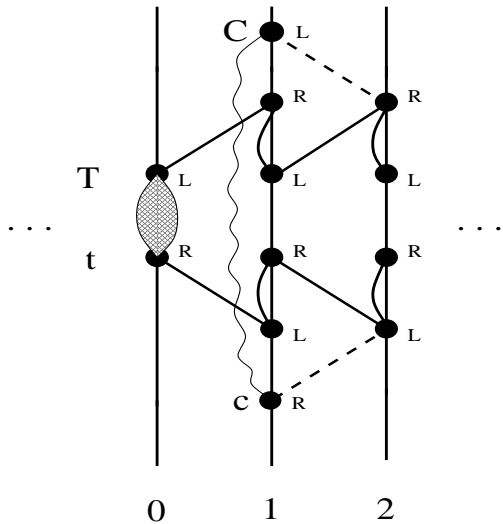}    
\vspace{1cm}    
\caption{The flavor mixing (dashed lines) between    
$\bar{C}_L\prod(\Phi/v) T_{R2}$ and $\bar{c}_R\prod(\Phi/v) t_{L2}$ cannot    
be rotated    
away by redefinitions of $T_{R2}$ and $t_{L2}$ without generating    
effective kinetic term mixing which leads    
to non-zero-mode flavor changing gluon vertices     
(in the broken phase where    
$\Phi\rightarrow v$; this mixing is actually a higher dimension    
operator).  The charm quark mass is thus generated when radiative    
corrections are included (wavy line).    
}      
\label{charm2}    
\end{figure}

Note that when we rotate away the direct charm mass terms on brane 1,    
we in general will obtain the linking mass terms,     
\be     
\ov{C}_L \left(\prod_a \Phi_{a,2} \right) T_{R2},\quad      
\ov{t}_{L2} \left(\prod_a \Phi_{a,2}^\dagger \right) c_R,\quad      
\ee     
where the $ \left(\prod_a \Phi_{a,i} \right) $ is a product    
of linking-Higgs, as shown in Fig.[5]. These terms can be viewed   
as mass terms,    
but in reality they are all higher dimension operators;    
since we are in the broken phase in which the    
$\Phi$'s all have VEV's we can only approximately    
describe these terms    
as though they are mass terms.       
However,     
even with the flavor mixing linking (higher dimension)    
mass terms and the direct mass terms, {\em at tree level}, (neglecting     
the gauge interactions on branes 1 and 2,) we can again    
perform a field redefinition as before and we     
still fail to generate a charm     
quark electroweak mass and only the top quark retains    
a nonzero electroweak mass.     
This     
can be seen readily    
as the EWSB condensate only couples to $T_{L0},\, t_{R0}$.     
We can rewrite $T_{L0},\, t_{R0}$ in terms of the     
redefined eigenstates of the electroweak preserving masses,    
\bear     
T_{L0}&=& \beta_3 T_L^{(0)} + \beta_2 C_L^{(0)} +\beta_1 U_L^{(0)}     
+ \mbox{heavy states} , \nonumber \\     
t_{R0}&=& \gamma_3 t_R^{(0)} + \gamma_2 c_R^{(0)} +\gamma_1 u_R^{(0)}     
+ \mbox{heavy states} .     
\eear     
After decoupling the heavy vector-like states, the $3\times 3$      
up-type quark mass matrix $M_U$     
\be     
{M_U}_{ij} \propto \beta_i \gamma_j,     
\ee     
is of rank 1.

However, the result of the field redefinition    
is that now {\em off-diagonal couplings to the gluons on    
branes 1 and 2 are generated!}    
When we now take into account the gauge      
interactions on branes 1 and 2, the charm quark does    
indeed      
obtain a nonzero mass from radiative corrections as shown in     
Fig.~\ref{charm2}. For this to occur we require the   
interference with the linking mass    
terms, because otherwise the gauge radiative corrections    
only produce multiplicative corrections to the (zero) mass    
on a given brane.

\begin{figure}[t]    
\vspace{5cm}    
\includegraphics{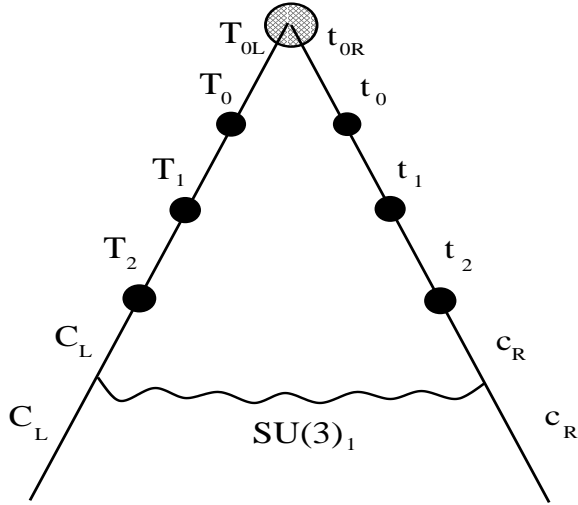}    
\vspace{1cm}    
\caption{The radiative correction diagram in the    
current eigenbasis for the induced charm mass.    
}      
\label{feyn1}    
\end{figure}

More explicitly, we now have the interbrane    
mass term  of the form, e.g., $ m'\ov{t}_{L2} c_R $ for    
the charm quark. This implies that    
on brane $2$ there is the overall mass term     
$ m'\ov{t}_{L2} c_R + M \ov{t}_{L2}t_{R2}$ where the second term    
the mass of the Dirac vectorlike $t$ quark.     
Thus, redefining $t_{R2} \rightarrow \cos\theta t_{R2}    
+\sin\theta c_R $ and $c_R \rightarrow -\sin\theta t_{R2}    
+\cos\theta  c_R $    
we can eliminate the direct charm quark mass term.    
However, in the kinetic terms we have:    
\be    
\bar{c}_R(i\slash{\partial} - \slash{A}_1)c_R     
+ \bar{t}_{R2}(i\slash{\partial} - \slash{A}_2)t_{R2}     
\ee     
Upon performing the redefinitions we generate off-diagonal    
transitions:     
\be    
\bar{c}_R(i\slash{\partial} - \slash{\tilde{A}}_1)c_R     
+ \bar{t}_{R2}(i\slash{\partial} - \slash{\tilde{A}}_2)t_{R2}     
+  \kappa(\bar{c}_R(\slash{A}_1-\slash{A}_2)t_{R2} + h.c.) + ...    
\ee    
where $\tilde{A}_{1(2)}=\cos^2\theta A_{1(2)}+\sin^2\theta A_{2(1)}$     
and $\kappa = \sin\theta\cos\theta$,    
and the ellipsis represents diagonal terms.    
On the left-handed side of Fig.[\ref{feyn1}] we    
also generate off-diagonal couplings of the form    
$ \kappa'(\bar{C}_L(\slash{A}_1-\slash{A}_2)T_{L2} + h.c.)$.    
In evaluating the induced charm quark mass and mixing it    
is useful to remain in the current eigenbasis in which    
the gluon interactions are diagonal.  We emphasize that this effect    
is different than that described by \cite{pomarol} in which    
localization produces off-diagonal flavor transitions amongst    
fermions coupled to KK mode vector bosons.    
    
\begin{figure}[t]    
\vspace{5cm}    
\includegraphics{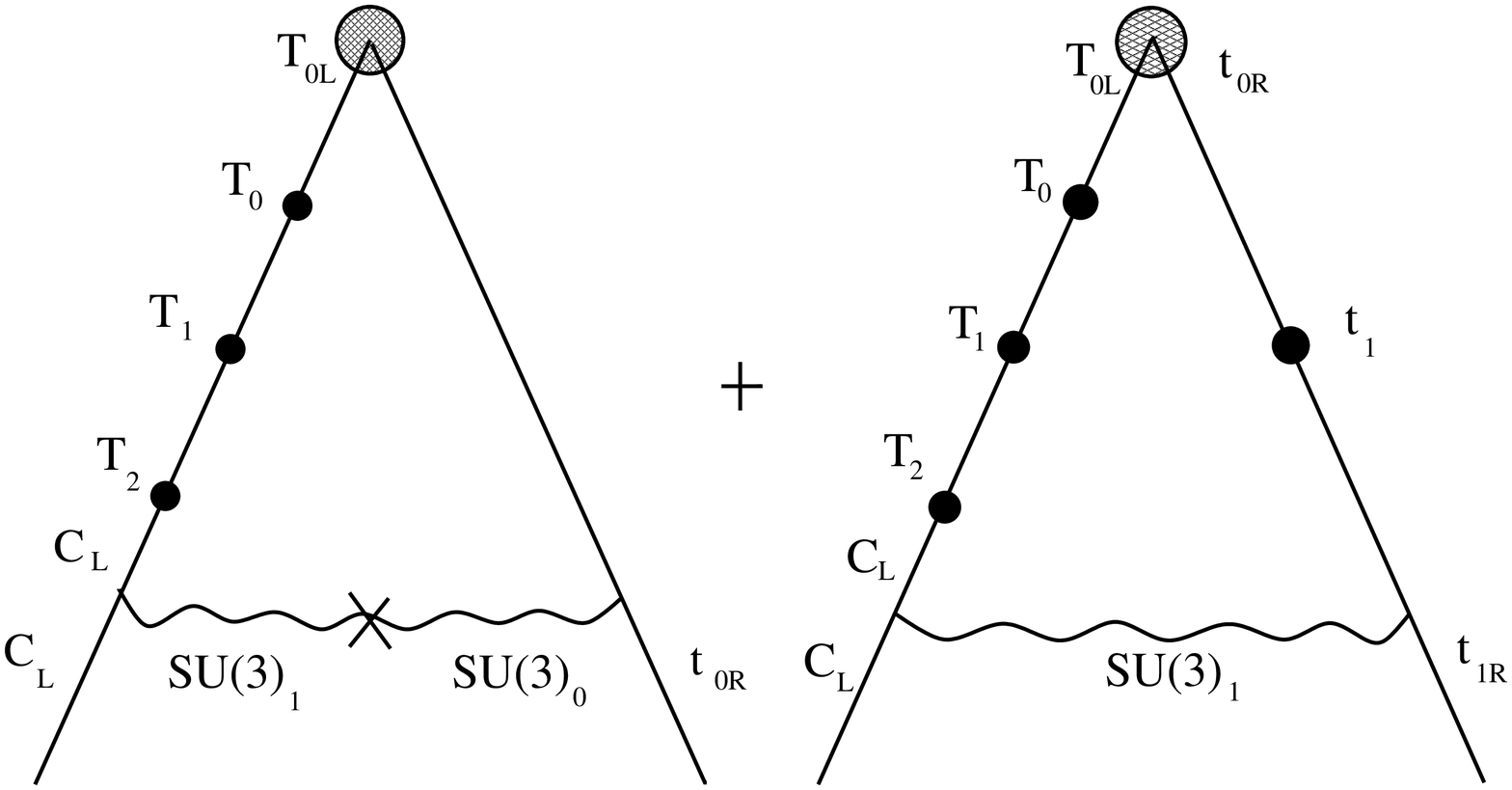}    
\vspace{1cm}    
\caption{The radiative correction diagrams in the    
current eigenbasis for the induced charm$_L$--top$_R$ mixing    
mass term.    
}      
\label{feyn2}    
\end{figure}

These off-diagonal kinetic terms, we emphasize, are higher dimension    
operators involving the link-Higgs fields! They take the apparent    
$d=4$ form only as a result of working in the broken phase of    
the $\Phi$'s. However, the result is that we have generated now    
an interaction that acts like {\em extended technicolor} \cite{eichten1}.    
When we include the radiative effects of the gluons we    
generate charm quark mass. In Fig.[\ref{feyn1}] we illustrate    
the diagram in the basis in which the gluon    
couplings are diagonal (the current eigenbasis)    
  We also generate radiative    
mixing between charm and top through diagrams as in      
Fig.[\ref{feyn2}] where now the mixing of the gluonic gauge groups    
on different branes must be    
included.    
    
\begin{figure}[t]    
\vspace{6cm}    
\includegraphics{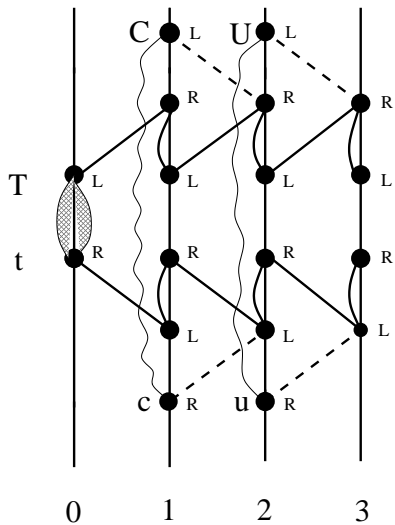}    
\vspace{1cm}    
\caption{The extension to include    
the up quark in a 4-brane model with    
radiatively generated mass and mixing.    
}      
\label{up1}    
\end{figure}    
    
The extension of the scheme to include the up quark    
mass generation and the mixing is shown in Fig.[\ref{up1}].    
Again, we require nearest neighbor mixing between branes    
which produces vanishing mass in tree approximation,    
but off-diagonal gluon vertices in the broken phase due to    
kinetic term mixing. The full mass    
matrix is regenerated when radiative corrections    
are included.

\begin{figure}[t]    
\vspace{6cm}    
\includegraphics{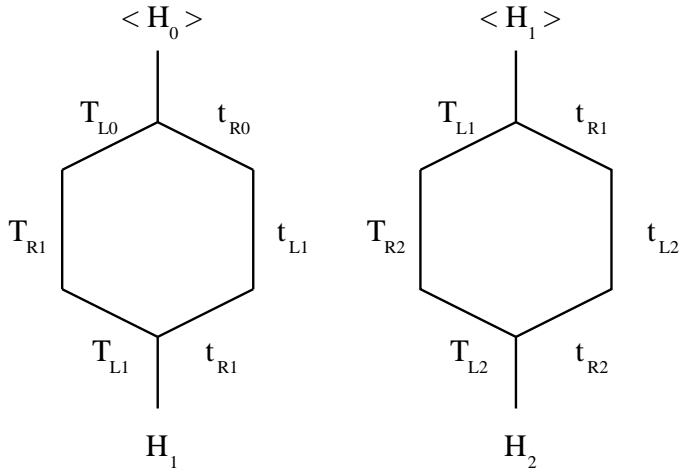}    
\vspace{1cm}    
\caption{The formation of composite Higgs fields on    
each brane and their propagation to subsequent branes. This sets    
up a tadpole on each brane which exponentially    
attenuates away from the brane 0 of the top    
condensate.    
}      
\label{higgs}    
\end{figure}

One can understand the origin of the mass matrix    
in the language of the ``shining Higgs VEV profile''    
as discussed     
in our previous paper~\cite{wang0}. The gauge interactions on branes 1--2     
are subcritical, so the Higgs bound states formed on these branes have     
positive squared masses. However, due to the links with brane 0, the      
composite Higgs fields on brane 1--2 will receive tadpole terms     
as shown in Fig.[\ref{higgs}], and therefore obtain nonzero VEVs.

{}From the shining and the flavor mixing effects, the final Higgs VEV     
will contain some small components of $\ov{C}_L c_R$ and $\ov{U}_L u_R$     
after diagonalization, which are responsible for generating the charm     
and up quark masses.     
    
To generate the down-type quark mass matrix requires    
a mechanism to first generate the $b$-quark mass.      
One possibility is to condense the $b$-quark     
as in the case of the top quark, and exploit    
a larger seesaw.  This encounters     
generally a large degree of fine-tuning; to have the    
large seesaw suppression of the physical $b$    
mass requires a larger vectorlike Dirac mass for the roaming    
b quarks, and this can turn off the    
condensate except for large supercritical coupling.    
An alternative and less fine-tuned approach is to    
exploit $SU(3)_0$ instantons on brane 0 which produce    
a 't Hooft determinant containing terms like    
$\sim \bar{b}_Lb_R\bar{t}_Lt_R + ...$   Then the nonzero    
$\VEV{\bar{t}t}$ induces the $b$-quark mass.  The magnitude    
of this condensate can be controlled by seesaw with the    
vectorlike $b$-quarks.  In any case, the dynamical    
and phenomenological details of the generation    
of the $b$ quark mass is a Top Seesaw modeling issue, and will    
be described in detail in a forthcoming paper by    
He, Hill and Tait \cite{he}.  For our    
present purposes we can simply assume that an induced    
$b$-quark mass or $\VEV{\bar{b}b}$ can be arranged for brane 0.    
The full model then takes the form of Fig.~\ref{up1}    
with $(u,c,t)$ replaced by $(d,s,b)$.  This produces a second species of Higgs    
boson, the $b$-Higgs $H_b$ which then shines through the    
bulk.    
    
Topcolor does not specifically address the issue    
of leptons.    
We can in principle use the $U(1)_{Y0}$ on the brane 0 to condense    
the $\tau$ lepton, and a corresponding seesaw to produce the physical    
$m_\tau$.  Alternatively, any new physics that produces the    
higher dimension operator $\bar{t}t\bar{\tau}\tau$ structure    
will suffice to give the $\tau$ lepton a mass.     
Having produced the  $\VEV{\bar{\tau}\tau}\neq 0$    
on brane 0, we again repeat the construction to provide     
the masses for $\mu$ and $e$. In the lepton case     
the $U(1)_Y$ radiative corrections replace the gluonic    
radiative corrections.  The neutrinos do not condense since    
$U(1)_Y$ does not produce a nontrivial `t~Hooft determinant (!),    
and we do not presently address the origin of the small neutrino    
Majorana masses.    
  
\subsection{Fourth Generation Condensates}  
  
A fourth generation scheme  
may have advantages  
for the lepton mass generation mechanism.  
Here we imagine condensing on a   
fourth generation brane$=0$ the three  
attractive channel condensates of $\VEV{\bar{T}T}$ and  
$\VEV{\bar{B}B}$ from $SU(3)_0$ and  
$\VEV{\bar{E}E}$ from a strong $U(1)_{Y0}$.  
The effective Higgs bosons are heavy,  
$\sim 1$ TeV.  
The model has an acceptably small positive  
$S\sim 2/3\pi$ and built in custodial $SU(2)$  
breaking from the $U(1)_{Y0}$ which may provide  
an acceptably large positive $T$ parameter contribution to bring  
the theory into the $S-T$ error ellipse.

 \begin{figure}[t]    
\vspace{7cm}    
\includegraphics{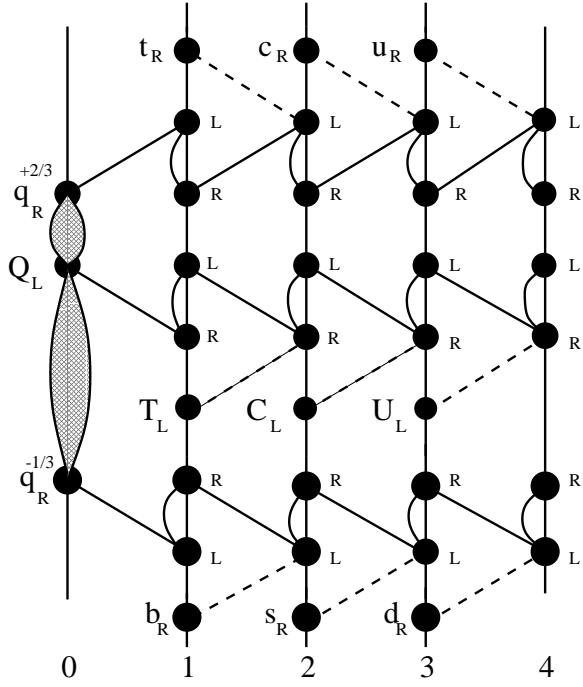}    
\vspace{1cm}    
\caption{The fourth generation condensate generating     
the up and down type quark masses.  We position the  
kinetic term partner $Q_R$ (which is nonchiral)  
next to the chiral zero mode $Q_L$.   
}      
\label{fourth}    
\end{figure}   
   
The structure of the quark sector for    
a 5-brane version of the model is shown in Fig.[\ref{fourth}].     
We use the strong $SU(3)_0$ (with strong    
$U(1)_{Y0}$ corrections) to form a $\VEV{\bar{B}B}    
\lta \VEV{\bar{T}T}$     
condensate on brane 0 of    
the fourth generation quarks. Thus, the $SU(3)$  
interaction overwhelms the tilting effect of the $U(1)_Y$.     
The quarks of the fourth generation roam through the bulk    
and propagate the composite Higgs.  The three lighter    
generations feel the condensate as in the Top Seesaw scheme    
of Section 3, as seen in Fig.[\ref{fourth}] which    
shows    
explicitly the quark sector.  The masses and    
CKM structure are then generated radiatively in analogy    
to the Top Seesaw scheme.  
    
\begin{figure}[t]    
\vspace{7cm}    
\includegraphics{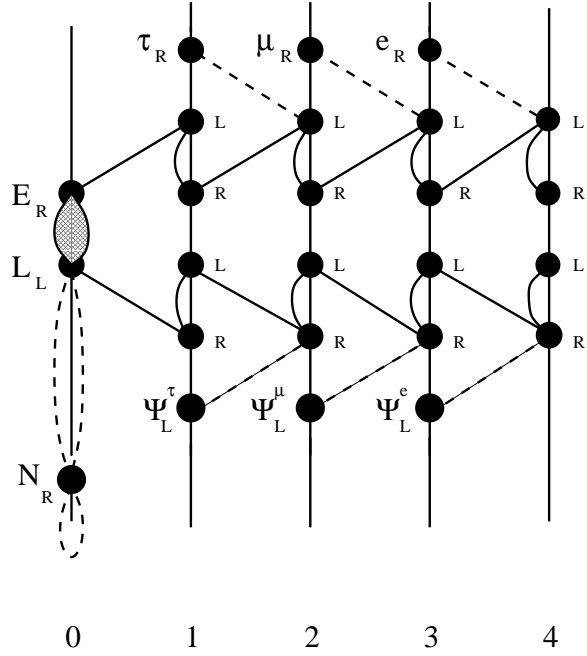}    
\vspace{1cm}    
\caption{The fourth generation condensate generating     
the lepton masses.  A single $N_R$ (which can be part  
of a vectorlike Dirac pair, and need not be chiral)  
is shown and an external mechanism gives it a Majorana mass.  
This is communicated to the left-handed neutrinos through  
the condensate with $L_L$.  
}      
\label{fourell}    
\end{figure}

In Fig.[\ref{fourell}] we illustrate how the lepton sector    
can be dynamically generated. Here there is a $U(1)_{Y0}$    
condensate of $\VEV{\bar{E}E}$ on brane 0 which     
produces the leptonic Higgs boson.  As before, the     
fermion masses are generated by linking flavor changing terms.    
    
We also include the right-handed neutrino $N_R$. We emphasize  
that this need not be chiral, i.e., it need not  
be a localized chiral zero mode associated with brane 0 kink.  
The $N_R$ is    
a gauge singlet so we can write down the Majorana mass terms,    
$\bar{N}_R^C N_{R}$ which presumeably comes from external  
physics, e.g., it may come from the effective Planck scale    
(we can certainly complicate the picture     
including all possible allowed Majorana   
and Dirac links within and between branes,    
 e.g. $\bar{N}_j^C N_{j+1}$, etc.).    
    
\begin{figure}[t]    
\vspace{6cm}    
\includegraphics{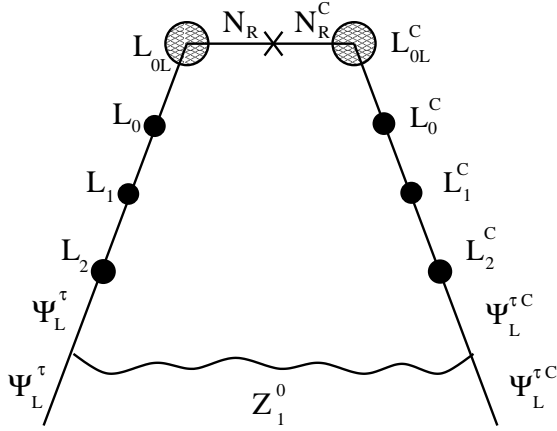}    
\vspace{1cm}    
\caption{The $\nu_\tau$ Majorana mass is  
radiatively induced by feeling the condensate  
and Majorana mass of $N_R$. This corresponds to  
an effective term $(\bar{\psi_L}\cdot H)^2$   
where $H$ is the Higgs through mixing with  
quarks and charged lepton, or the neutrino condensate effective  
boundstate Higgs $\sim \bar{N}_RL_L$.  
}      
\label{feyn5}    
\end{figure}

To produce small Majorana mass terms for the known neutrinos     
we first require a mechanism to generate a Dirac mass  
or condensate  
for the 4th generation neutrino, $\sim M\bar{\nu}_L N_R$  
on brane 0.  This requires one of two  
options: (a) A condensate such  
as $\VEV{\bar{L}N_R}\neq 0$ must form involving yet a new  
strong gauge group, such as $U(1)_{B-L}$ {\em or}  
(b) A higher dimension operator exists which allows   
the bilinear $\bar{L}N_R$ to feel the electroweak condensates,  
as in:  
\be    
\frac{1}{M^2}\bar{L}^a_LN_R(\bar{B}_R T^a_L)\qquad    
\frac{1}{M^2}\bar{L}^a_LN_R(\bar{E}_R L_L)^C   
\ee   
Once the master 4th generation  
Dirac mass is established we can invoke the   
seesaw.  We depict this in Fig.[\ref{fourell}].     
The Feynman diagram of   Fig.[\ref{feyn5}]  
then shows the formation of the radiatively induced Majorana  
mass (by, e.g., $Z$ exchange) for the $\nu_\tau$.  
Similar mixings and mass terms arise for the first and  
second generation neutrinos in analogy to the quark and   
charged lepton masses.  
  
We mention that one should be wary of the possibility of enhanced    
proton decay coming from the 't Hooft process with the strong    
$SU(2)_L$ gauge groups located on various branes.  This is an    
issue for ``topflavor'' models which we defer to another session.    
Moreover, as discussed in Ref.~\cite{pomarol}, the KK gauge bosons can induce     
flavor-changing effects in the split fermion generation models. This     
puts a strong constraint on the KK gauge boson masses. In our model,     
the first two generations are localized away from the EWSB brane.     
Flavor-changing effects involving the first two generations from     
heavy gauge boson exchanges can be suppressed if the link VEVs     
associated with the first two generation branes are much larger     
than the weak scale (since they are not directly related to EWSB).     
We have not yet touched upon the vacuum alignment of the   
VEV's of the composite Higgs, possibly through mechanisms such as radiative  
corrections,    
higher order couplings. Such questions have been discussed by   
earlier works on dynamical electroweak symetry breaking, e.g. \cite{he},   
we will put off discussion on our specific model till the future work  
\cite{wang5}.

\section{Discussion and Conclusion}

In conclusion, we have given a description       
of a fairly complete extension of the  
Standard Model with dynamical electroweak symmetry breaking.  
This arises from the bulk $1+4$ dimensions  
as a $1+3$ dimensional effective theory after  
remodeling.        
We have kept only a small number of lattice  
slices (branes) as a minimal approximation with       
thinned degree of freedom.       
            
A dynamical       
electroweak symmetry breaking scheme emerges   
naturally in this       
description, as first anticipated  
by Dobrescu \cite{dobrescu2}.  We see immediately       
the emergence of an imbedding       
of QCD as in $SU(3)\rightarrow SU(3)_1\times SU(3)_2 ...$,   
and the appearance of vectorlike partners  
of the elementary fermions such as the top quark.    
With fermionic  
localization we can have flavor dependent  
couplings to these gauge groups, and trigger  
the formation of condensates using   
localization background fields or warped geometry.  
      
These elements are all part of the structure       
of Topcolor, \cite{hill0},   
and the Top Seesaw \cite{topseesaw},  
and we are thus led naturally  
to this class of extra-dimensional        
models in which the electroweak symmetry       
is broken dynamically.  However, one can go beyond  
these schemes to, e.g., a fourth generation  
scheme which is somewaht more  
remniscent of Technicolor and may have  
direct advantages for the lepton sector masses.     
       
 One can always discard       
the notion of extra-dimensions and view this       
as an extension of the Standard Model       
within $1+3$ dimensions with extra discrete symmetries,  
however the specific structures we have considered are  
almost compelled by extra dimensions.       
The connection to extra dimensions is made through   
{\em remodeling} \cite{wang0, wang1}, bulk  
inhabitation of gauge fields \cite{dienes}, the      
the transverse lattice \cite{trans}, hidden   
local symmetries \cite{hidden},  
etc., and may be viewed as a        
manifestly gauge invariant       
low energy effective theory for an      
extension of the Standard Model in $1+4$.

Remodeling is a remarkable model building tool,  
 and a system of new organizational principles.  
Remodeling has guided our thinking in producing   
the present sketch of a full theory  
of flavor physics based upon Top Seesaw, something which has  
not been previously done.    
Much work remains to sort out and to  
check that the systematics of experimental constraints can  
be accomodated \cite{he,wang5}, and to see if the model survives as a natural  
scheme without a great deal of fine-tuning.  It is already  
encouraging that the Top Seesaw model is a strong dynamics  
that is consistent with experimental $S-T$ constraints.

\section*{Acknowledgements}       
       
We wish to thank W. Bardeen for useful discussions.       
H.-C. Cheng is supported by the Robert R. McCormick Fellowship and by       
DOE grant DE-FG02-90ER-40560.        
Research by CTH and JW was supported by the U.S.~Department of Energy       
Grant DE-AC02-76CHO3000.       
  
\newpage    
\setcounter{equation}{0}
\renewcommand{\theequation}{A.\arabic{equation}}
\section*{Appendix A: Three Brane Example of Gauge Fields sans     
Translational Invariance}    
    
We now wish to address the impact of    
the breaking of translational invariance in $x^5$    
on the physics of the effective $1+3$ Lagrangian.    
It is generally advantageous    
to thin the degrees of freedom in the     
lattice description  of the extra dimensions.    
We can construct a coarse grain $n$-brane model with    
$n << N$  as a crude approximation    
to a fine grained $N$-brane model. Such a description    
can be improved in principle by a block-spin renormalization    
group, which is beyond the scope of our present    
discussion.     
    
Consider, for example,    
a $3$-brane model.    
The effective $1+3$    
Lagrangian now contains 3 copies of the Standard Model     
gauge group and link fields interpolating      
each of the $SU(3)_C$, $SU(2)_W$, and $U(1)_Y$     
groups in the aliphatic configuration.      
The pure gauge Lagrangian in $1+3$ dimensions     
for $3$ copies of QCD is given by:     
\be     
{\cal{L}}_{QCD}= -\frac{1}{4}\sum_{j=0}^2 G_{i\mu\nu}^B G_j^{B\mu\nu} +     
\sum_{j=1}^{2} D_{\mu}\Phi_j^\dagger D^{\mu}\Phi_j.      
\ee      
where $G^B_{i\mu\nu}$ has been rescaled so      
that the gauge coupling $\tg_{3,j}$      
appears in the covariant derivative.      
The electroweak gauge Lagrangian can be     
written down analogously.     
     
After substituting the VEVs of the link fields,     
\be       
\Phi_j \rightarrow v_j\exp(i\phi^B_j T^B_j /v),     
\ee     
the $\Phi_j$ kinetic terms lead to a mass-squared matrix for the gauge     
fields:     
\be     
\sum_{j=1}^{2} \half v_i^2(\tg_{3,(i-1)}A^B_{(i-1)\mu} -\tg_{3,i}A^B_{i\mu})^2      
\ee      
This mass-squared matrix can be written as an $3\times 3$      
matrix sandwiched      
between the column vector $A=(A^B_{0\mu}, A^B_{1\mu},       
A^B_{2\mu})$, and it's transpose, as $A^TMA$, where:      
\be      
\label{mat1}      
M =   \half \left(      
\begin{array}{ccc}      
(\tg_{3,0})^2 v_1^2 &-(\tg_{3,0}\tg_{3,1})v_1^2 &0 \\      
-(\tg_{3,0}\tg_{3,1})v_1^2 & (\tg_{3,1})^2 (v_1^2+v_2^2)       
&-(\tg_{3,1}\tg_{3,2})v_2^2  \\      
0& -(\tg_{3,1}\tg_{3,2})v_2^2  &(\tg_{3,2})^2 v_2^2 \\      
\end{array} \right).      
\ee      
 where we have kept the full set of effects    
of $j$-dependence in $v_j$ and $\tg_{3,j}$.    
      
We can diagonalize the mass-matrix as:       
\be       
A_{j\mu} = \sum_{n=0}^{2} a_{jn} \tilde{A}_{\mu}^n.       
\label{AA}      
\ee      
The $a_{jn}$ form a normalized eigenvector ($\vec{a}_{n}$)      
associated with the $n$th eigenvalue.      
The eigenvectors and the corresponding eigenstate masses      
for common $\tg_3$ and $v$, which corresponds to the flat extra dimension case,      
were obtained in the previous papers~\cite{wang0,wang1},      
\bear      
\vec{a}_0 &=& \frac{1}{\sqrt{3}}(1,\, 1,\, 1) \nonumber \\      
\vec{a}_1 &=& \sqrt{\frac{2}{3}} \left( \cos\frac{\pi}{6},\,       
\cos\frac{3\pi}{6},\,      
\cos\frac{5\pi}{6} \right) = \frac{1}{\sqrt{2}} (1,\, 0,\, -1) \nonumber \\      
\vec{a}_2 &=& \sqrt{\frac{2}{3}} \left( \cos\frac{2\pi}{6},\,       
\cos\frac{6\pi}{6},\,      
\cos\frac{10\pi}{6} \right) = \frac{1}{\sqrt{6}} (1,\, -2,\, 1)      
\eear      
\be      
(M_0,\, M_1,\, M_2)\; =\; 2\tg_3 v\:\left(0,\, \sin\frac{\pi}{6},\,       
\sin\frac{\pi}{3}      
\right) \; =\;  \tg_3 v\: (0, \, 1,\, \sqrt{3}),      
\ee      
      
The expressions for the eigenvectors and eigenvalues for general       
$\tg_{3,i}$ and $v_i$      
are more complicated. However, if $\tg_{3,0}v_1,\, \tg_{3,1} v_1\,       
\gg \,\tg_{3,1} v_2,\,      
\tg_{3,2} v_2$, we have a sequential    
decoupling. In this case    
the $SU(3)_0\times SU(3)_1$ is first broken down to the       
diagonal      
$SU(3)'$ by $\Phi_1$, then $SU(3)'\times SU(3)_2$ is broken by $\Phi_2$ to       
$SU(3)_{QCD}$      
at a lower scale. In this case, the eigenstates and their masses are      
given approximately by:      
\bear      
\vec{a}_2 &\approx &\frac{1}{\sqrt{\tg_{3,0}^2 + \tg_{3,1}^2}}\,      
( \tg_{3,0},\; -\tg_{3,1},\; 0), \nonumber \\      
\vec{a}_1 &\approx & \frac{1}{\sqrt{(\tg_{3,0}^2\tg_{3,1}^2      
+\tg_{3,1}^2\tg_{3,2}^2+\tg_{3,0}^2\tg_{3,2}^2)      
(\tg_{3,0}^2 + \tg_{3,1}^2)}}\, (\tg_{3,0}\tg_{3,1}^2,\; \tg_{3,0}^2\tg_{3,1},      
\; \tg_{3,2}\,(\tg_{3,0}^2+\tg_{3,1}^2)), \nonumber \\      
\vec{a}_0 &=& \frac{1}{\sqrt{\tg_{3,0}^2\tg_{3,1}^2      
+\tg_{3,1}^2\tg_{3,2}^2+\tg_{3,0}^2\tg_{3,2}^2}}\, (\tg_{3,1}\tg_{3,2},\;      
\tg_{3,0}\tg_{3,2},\; \tg_{3,0}\tg_{3,1}),  \\      
M_2^2 &\approx & (\tg_{3,0}^2+\tg_{3,1}^2)\, v_1^2, \;\;\;\;      
M_1^2 \approx \frac{\tg_{3,0}^2\tg_{3,1}^2      
+\tg_{3,1}^2\tg_{3,2}^2+\tg_{3,0}^2\tg_{3,2}^2}{\tg_{3,0}^2+\tg_{3,1}^2}\,       
v_2^2, \;\;\;\; M_0^2=0.      
\eear       
Note that unlike the translationally invariant case, the massive states  
do not necessarily correspond to the lowest eigenstates in the continuum  
limit.

\newpage    
\setcounter{equation}{0}
\renewcommand{\theequation}{B.\arabic{equation}}
\section*{Appendix B: Chiral Fermions and a Discretized Version     
of the Jackiw-Rebbi Domain Wall}

In $1+4$ dimensions free fermions are vectorlike.      
Chiral fermion zero modes can be obtained by        
using domain wall kinks in a background      
field  which couples to  the fermion       
like a mass term. This can trap a chiral zero-mode      
at the kink~\cite{jackiw}. This mechanism can be generalized to       
the lattice action~\cite{kaplan}. We now discuss the chiral fermions      
in the discretized version of the Jackiw-Rebbi domain wall.      
      
We first consider an infinite fifth dimension, (i.e., there are infinite      
number of $SU(3)$'s for QCD,) and for simplicity, we assume that       
$\tg$ and $v$ are constant. From eq.(\ref{warped_fermion}) we see that       
the kinetic term in the fifth dimension appears as a      
fermion mass terms on the lattice:      
\be   
i\overline{\Psi} \gamma^5\partial_5 \Psi    \sim  
\tg v\, \overline{\Psi}_{iL} \Psi_{iR}       
-\tg v\, \overline{\Psi}_{iL} \Psi_{(i-1)R} + h.c..      
\ee    
Note that the derivative hops $[i,L]\rightarrow [(i-1),R]$.  
We can equally well represent the derivative as:  
\be   
i\overline{\Psi} \gamma^5\partial_5 \Psi    \sim  
\tg v\, \overline{\Psi}_{iL} \Psi_{(i+1)R}       
-\tg v\, \overline{\Psi}_{iL} \Psi_{iR} + h.c..      
\ee    
hopping  $[(i-1),L]\rightarrow [i,R]$.  
  
The mass matrix between $\overline{\Psi}_L$ and $\Psi_R$,      
$\overline{\Psi}_L\, M_f \,\Psi_R$ in the first  
convention is:      
\be      
M_f = \tg v \left(      
\begin{array}{cccccc}      
\ddots & \ddots & \cdots & & & \\  
\ddots & 1 & 0 & \cdots & & \\      
\cdots & -1 & 1 & 0  & \cdots &  \\      
&  0   & -1 & 1 &  0 & \cdots  \\      
& \vdots & 0 & -1  & 1 & \ddots  \\      
& & \vdots & \vdots & \ddots & \ddots   \\  
 \end{array}      
\right).      
\ee      
A left-handed chiral zero mode can be localized at $y=y_k$ by       
a kink fermion mass term which has $m_{\Psi}(y<y_k)>0$ and       
$m_{\Psi}(y>y_k)<0$. In the discrete version,   
one can add positive and negative masses to the diagonal term,       
$-m \ov{\Psi}_{iL} \Psi_{iR}$ for $i<k$ and $i>k$ respectively  
as in Ref.~\cite{kaplan}.  
For example, with the kink at $k=3$ we have:  
\be      
M_f =  \left(      
\begin{array}{ccccc}      
\tg v +m & 0 & \cdots & & \\      
-\tg v & \tg v+m & 0  & \cdots &\\      
   0   & -\tg v & \tg v &  0 & \cdots  \\      
\vdots & 0 & -\tg v  & \tg v-m & \ddots \\      
 & \vdots & \ddots & \ddots & \ddots      
\end{array}      
\right).      
\ee    
The {\em approximate} solution for a zero mode is then:  
\be      
\psi_L \propto \sum_i\epsilon^{|i-k-1/2|}\, \Psi_{iL}, \quad \quad      
\epsilon \approx \frac{\tg v}{\tg v+ m} < 1,      
\ee        
where, in this case we require $m<< \tg v$   
and hence localization of the zero mode requires   
a fine grain lattice such that      
$|\tg v -m| < \tg v$.  
  
Alternatively, and more efficient for a  
coarse grain lattice,   
we can give a positive      
mass to the diagonal mass term $m \ov{\Psi}_{iL} \Psi_{iR}, \; m>0$      
for $i<k$ and a negative mass to the off-diagonal mass term      
$-m \ov{\Psi}_{iL} \Phi_{i} \Psi_{(i-1)R}/v$ for $i>k$.  
As in the previous example, we now have:  
\be      
M_f =  \left(      
\begin{array}{ccccc}      
\tg v +m & 0 & \cdots & & \\      
-\tg v & \tg v+m & 0  & \cdots &\\      
   0   & -\tg v & \tg v &  0 & \cdots  \\      
\vdots & 0 & -\tg v -m  & \tg v & \ddots \\      
 & \vdots & \ddots & \ddots & \ddots      
\end{array}      
\right).      
\ee   
This enhances      
the diagonal links for $i<k$ and the off-diagonal links for $i>k$.      
A left-handed chiral zero mode then arises centered around $\Psi_{kL}$      
which has the ``weakest links''. One can easily check that the state      
\be      
\psi_L \propto \sum_i \epsilon^{|i-k|}\, \Psi_{iL}, \quad \quad      
\epsilon = \frac{\tg v}{\tg v+ m} < 1,      
\ee      
is a zero mode, while there is no normalizable right-handed zero mode.      
The width of the zero mode becomes narrower for smaller $\epsilon$.      
In the limit $m \gg \tg v$, the zero mode is effectively localized      
only on the lattice point $k$. Similarly, a right-handed chiral mode      
can be localized by considering the opposite mass profile,  
and the method can easily be adapted to the opposite   
derivative (hopping) definition.     
      
If we compactify the extra dimension with the periodic boundary condition,      
there will be another zero mode with the opposite chirality localized      
at the anti-kink of the mass term. In general, the pair of zero modes      
will receive a small mass due to the tunneling between the finite distance      
of the kink--anti-kink separation unless some fine-tuning is made.      
With the $S^1/Z_2$ orbifold compactification, however, one of the zero      
mode will be projected out. One can see that in the discrete aliphatic      
model, the boundary conditions removes one chiral fermion at the end      
of the lattice point, so there must be a chiral fermion left massless      
due to the mismatch of the numbers of the left-handed and right-handed      
fermions. The massless chiral fermion can be localized anywhere on the       
lattice using the discrete domain wall mass term.

\vspace*{1.0cm}       
     
\frenchspacing       
       
\end{document}